\documentstyle[aps,prd]{revtex}
\input epsf 
\tightenlines
\begin{document}

\begin{titlepage}
\preprint{MADPH-99-1108}

\title{A note on the computation of mode sums}

\author{Theodore J. Allen} 
\address{Department of Physics, Eaton Hall \\
Hobart and William Smith Colleges \\
Geneva, New York 14456 USA }
\author{M. G. Olsson}
\address{Department of Physics, University of Wisconsin, \\
1150 University Avenue, Madison, Wisconsin 53706 USA }

\author{Jeffrey R. Schmidt}
\address{Department of Physics, University of Wisconsin-Parkside\\
900 Wood Road,
Kenosha, Wisconsin  53141 USA}

\date{\today}
\maketitle
\thispagestyle{empty}

\begin{abstract}
The computation of mode sums of the types encountered in basic quantum
field theoretic applications is addressed with an emphasis on their
expansions into functions of distance that can be interpreted as
potentials.  We show how to regularize and calculate the Casimir energy for
the continuum Nambu-Goto string with massive ends as well as for the
discrete Isgur-Paton non-relativistic string with massive ends. As an
additional example, we examine the effect on the interquark potential of a
constant Kalb-Ramond field strength interacting with a QCD string.
\end{abstract}

\end{titlepage}

\section{Introduction}\label{sec:intro}

The determination of the Casimir, or zero-point, energy for a quantum
system through the evaluation of the zero mode energy sum is a problem of
quite practical importance.  Casimir originally considered the effect of
conducting plates on the QED vacuum energy due to the changed boundary
conditions\cite{casimir}.  A review of the Casimir effect can be found in
the Physics Report by Plunien, M\"uller, and Greiner\cite{plunien:1986ca}.

The zero-point energy sums are almost always divergent and need
regularization in order to be interpreted sensibly.  The resulting Casimir
energy may be interpreted as a background energy or potential.  This is
made most evident if the sum can be expressed in terms of simple functions
of distance, which might be put into one-to-one correspondence with simple
potential terms such as Coulombic or centripetal potential components.  The
purpose of this paper is to illustrate how this can be done for several
classes of mode sums encountered recently in the literature.

We begin in section \ref{sec:lambiasenesterenko} by examining the
computation of the Casimir energy for a static string with massive ends
that is allowed to vibrate transversely.  This problem was solved in closed
integral form by Lambiase and Nesterenko \cite{lambiasenesterenko}.  We
review the derivation of the integral form and then find an approximate
series representation for large quark end masses.  

In section \ref{sec:Besselsum}, we find useful series approximation for a
sum of Bessel functions that appears throughout this paper, and elsewhere
in physics literature.

In section \ref{sec:monopole} we examine the effect of a particular
background Kalb-Ramond field upon the Casimir energy.  The Kalb-Ramond
field we examine is one giving rise to a constant three-form field
strength.  The Kalb-Ramond field strength may be viewed as the dual of the
abelian monopole current found in lattice QCD that likely leads to the
formation of the chromoelectric flux tubes.  The effect on the Casimir
energy of string self-interaction mediated by the Kalb-Ramond field and the
dilaton is to renormalize the string tension \cite{elizaldeodintsov}, which
also happens in a background Kalb-Ramond field.  We find additional non-linear
corrections to the static potential in a background Kalb-Ramond field.

Finally, we examine the Casimir energies in the discretized string model of
Isgur and Paton \cite{isgurpaton}.  We find the interesting result that
although the Casimir energies of the continuum and discrete strings agree
for fixed-end strings, the finite quark mass corrections are of a completely
different form.

\section{Continuum QCD string with massive ends}\label{sec:lambiasenesterenko}

\begin{figure}[htbp]
\epsfxsize = \textwidth \epsfbox{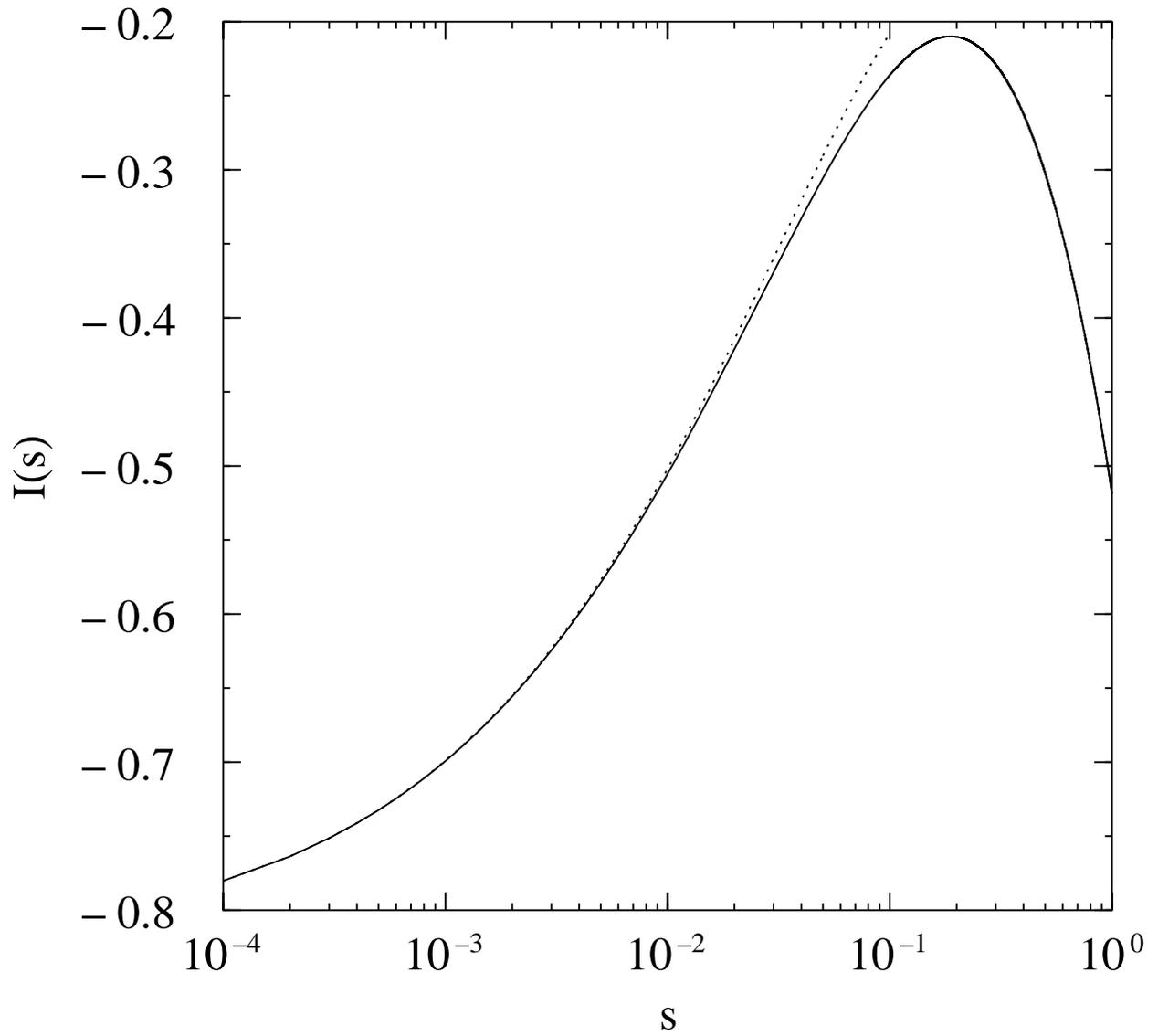}
\vskip 1 cm
\caption{Comparison of exact result with series representation of Casimir
energies of string with massive ends. Numerical evaluations of
Eq.~(\protect\ref{9}), dotted line, and Eq.~(\protect\ref{23}), solid
curve, for small values of $s$.}
\label{fig:one}
\end{figure}

\begin{figure}[htbp]
\epsfxsize = \textwidth \epsfbox{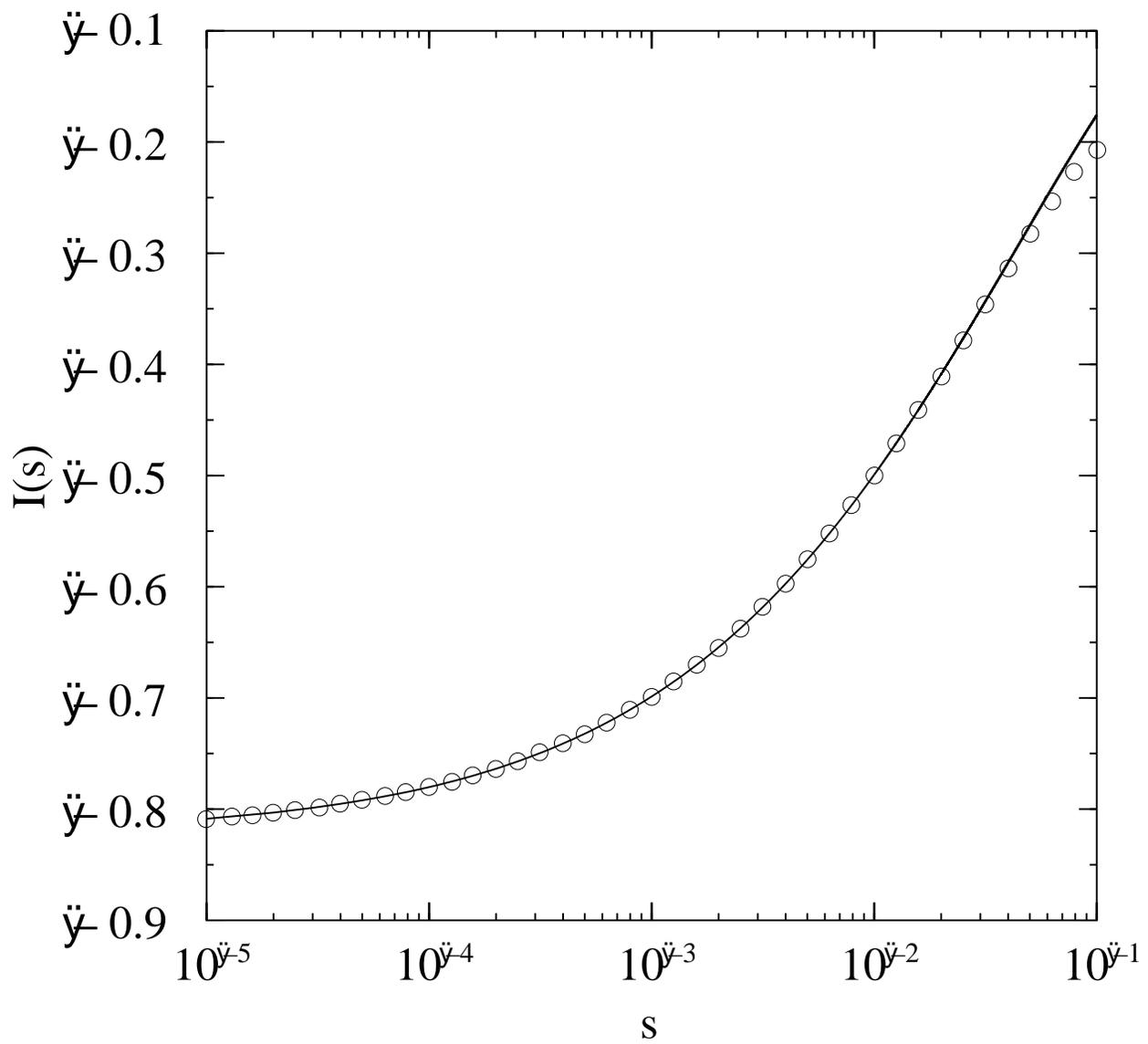}
\vskip 1 cm
\caption{Comparison of sum of Bessel functions and series expansion. Numerical evaluations of Eq.~(\protect\ref{Ksum}),
circles, and Eq.~(\protect\ref{23}), solid curve, for small values of $s$.}
\label{fig:two}
\end{figure}

The eigenfrequencies for a vibrating Nambu-Goto string of length $r$ and
tension $a$, terminated by two massive quarks of mass $M$ that are allowed
to move only transversely has been given by Lambiase and Nesterenko
\cite{lambiasenesterenko} as the roots of the transcendental equation
\begin{equation} 
f(\omega)=(M^2\omega^2-a^2)\sin\omega r - 2Ma \omega \cos\omega r=0 \ .
\end{equation} 
The Casimir energy, which is half the sum of these eigenfrequencies, can be
calculated with aid of the Watson-Sommerfeld transform
\begin{equation} \label{CasDef}
E_{\rm Casimir} = {1\over 2}\sum_n \omega_n  =  {1\over 4\pi
i} \oint_C \omega \, d\omega\, {d\over d\omega}\ln f(\omega) \ ,
\end{equation} 
where the contour $C$ encloses the roots of $f(\omega)$, all of which lie
on the positive real axis for physically meaningful frequencies.  The
energy computed in Eq.~(\ref{CasDef}) is the energy per transverse degree
of freedom.  In $D$ dimensions, there would be an extra factor of $D-2$.
The sum in Eq.~(\ref{CasDef}) is formally infinite and must be regularized
and renormalized.
We deform the contour to include the entire left half-plane and
keep only the regular portion
\begin{equation}
{1\over 2}\sum_n \omega_n^{\,\rm reg} = -\, {1\over 2\pi } \int_{0}^{\infty}
y \, dy\, {d\over dy}\ln\left[(M^2 y^2+a^2)\sinh ry +2Ma y\cosh ry\right] \
.
\end{equation}
We renormalize by subtracting off the part of this integral for which
$r\rightarrow \infty$, to arrive at
\begin{equation}
{1\over 2}\sum_n \omega_n^{\,\rm reg, ren}  = -\,{1\over 2\pi }
\int_{0}^{\infty} y \, dy \, {d\over dy}\ln{[(M^2
y^2+a^2)\sinh ry +2Ma y\cosh ry]\over {1\over
2}e^{ry}[M^2y^2+a^2+2May]} \ ,
\end{equation}
which we simplify to
\begin{equation}
{1\over 2}\sum_n \omega_n^{\,\rm reg, ren} = -\,{1\over 2
\pi}\int_0^{\infty} y\, dy\, {d\over dy}\ln\left[1-e^{-2ry}\left({My-a\over
My+a}\right)^2\right] \ ,
\end{equation}
and integrate by parts to arrive at
\begin{eqnarray}
{1\over 2}\sum_n \omega_n^{\,\rm reg, ren} &=& {1\over 2 \pi}\int_0^{\infty}
dy\, \ln\left[1-e^{-2ry}\left({My-a\over My+a}\right)^2\right] \\ &=&
{1\over 2 \pi r}\int_0^{\infty} dy \,\ln\left[1-e^{-2y}\left({y-s\over
y+s}\right)^2\right] \ ,
\end{eqnarray}
with the dimensionless quantity
\begin{equation}\label{sdef}
s={ar\over M} \ . 
\end{equation}
Defining the quantity $I(s)$, we have the exact result
\cite{lambiasenesterenko}
\begin{equation}\label{9}
I(s)= 2 \pi r E_{\rm Casimir} = \int_0^{\infty} dy
\,\ln\left[1-e^{-2y}\left({y-s\over y+s}\right)^2\right] \ . 
\end{equation}
The evaluation of the Casimir energy for small $s$ can be performed as follows.
For very small $s$ (large masses $M$) the approximation,
\begin{equation}\label{10} 
\left({x-s\over x+s}\right)^{2n}\approx e^{-\,{4ns\over x}}
e^{-\,{4ns^3\over 3 x^3}\,+\,\cdots} \ ,
\end{equation}
is extremely accurate. The logarithm can be expanded
\begin{eqnarray}
I(s) &=& - \, \sum_{n=1}^{\infty}{1\over n}\int_0^{\infty}
\left({x-s\over x+s}\right)^{2n}e^{-2nx}\, dx \\
&\approx& -\sum_{n=1}^{\infty}{1\over n}\int_0^{\infty}
e^{-\,2nx\,-\,{4ns\over x}}\, dx \ ,
\end{eqnarray}
and used with the integral representation of the modified Bessel function
\begin{equation} 
\int_0^{\infty} x^{\nu-1} e^{-\,{\beta\over x}\, -\, \gamma x} \, dx =
2\left({\beta\over\gamma}\right)^{{\nu\over 2}}
K_{\nu}(2\sqrt{\beta\gamma}) \ ,
\end{equation}
to obtain
\begin{equation} \label{Ksum}
I(s) \approx -\,\sqrt{8s} \sum_{n=1}^{\infty} {1 \over n} K_1(2n\,\sqrt{8s}) 
\end{equation}
as an excellent approximation for $s$ small.  Using the results of 
section \ref{sec:Besselsum} below, we find
\begin{eqnarray}\label{23}
I(s) &=&-\,{\pi^2\over 12} + \sqrt{2\pi^2 s} + ({4\gamma - 2})s -
4s\,\sinh^{-1}\left({\pi\over\sqrt{8s}}\right) + \cdots \ ,
\end{eqnarray}
which agrees extremely well with numerical evaluations of the integral for
small $s$, as can be seen in Fig.~\ref{fig:one}. A comparison of the series
(\ref{23}) and the sum (\ref{Ksum}) is shown in Fig.~\ref{fig:two}. If more
accuracy is needed for larger values of $s$, higher terms in $s$ in the
exponent of Eq.~(\ref{10}) can be retained and those contributions to the
exponent expanded in a power series.

If the quark mass $M$ is large, we can use the approximation
Eq.~(\ref{23}), to calculate the lowest order corrections in $1\over M$ to
the Casimir energy.  From the definition Eq.~(\ref{9}), we find the Casimir
energy per transverse degree of freedom
\begin{equation} \label{continCasimir}
E_{\rm Casimir} = -\,{\pi \over 24 r} + \sqrt{{a\over 2Mr}} + {\cal
O}({a\over M}) \ . 
\end{equation}

\section{Evaluation of a sum of Bessel functions}\label{sec:Besselsum}

In this section we evaluate the sum
\begin{equation}
G(x) = \sum_{n=1}^{\infty} {1 \over n} K_1(nx) \ ,\label{eq:Ksum}
\end{equation}
for small $x$.  This sum appears in the evaluation of functional
determinants \cite{dowker} as well as Casimir energies
\cite{lambiasenesterenko,elizalderomeo,nesterenkopirozhenko}. Related sums
of Bessel functions appear in generic one-loop finite temperature effective
potentials for GUTs \cite{easthermoreau}, and also appear in the
mathematics literature \cite{watson}.

Evaluation of this sum using zeta functions quickly becomes problematic
because no matter how small $x$ is, for sufficiently large $n$ values,
$n\,x\rightarrow \infty$ and small $x$ expansions of $K_1(x)$ cannot be
used. However, we can use another integral representation of the Bessel
function,
\begin{equation} 
K_1(nx) = {2 \pi x\over n}\int_0^{\infty} {\cos(2 \pi nt)\, dt\over ((2\pi
t)^2+x^2)^{3\over 2}} \ ,
\end{equation}
together with the sum
\begin{equation}
\sum_{n=1}^{\infty} {\cos(2 \pi nt)\over n^2} = \pi^2 B_2(t - [t]) \ ,
\end{equation}
where $[t]$ is the greatest integer less than $t$ and $B_2(t)$ is the
second Bernoulli polynomial,
\begin{equation}
B_2(t) = {1\over 6}\,-\,{t}\ +\ {t^2} \ .
\end{equation}
We obtain
\begin{eqnarray}
\sum_{n=1}^{\infty} {K_1(nx)\over n}&=&2\pi x\int_0^{\infty}
{dt\over ((2\pi t)^2+x^2)^{3\over 2}}\left(\sum_{n=1}^{\infty} 
{\cos(2\pi nt)\over n^2}\right) \nonumber \\
&=&2 \pi x\int_0^{\infty} {\pi^2B_2(t - [t])\ dt\over ((2\pi t)^2+x^2)^{3\over
2}} \  .
\end{eqnarray}

We may split this expression into the first term and a remainder
\begin{equation}
G(x) = 2\pi x \int_0^1 {\pi^2 B_2(t)\over (x^2+(2\pi t)^2)^{{3\over 2}}}
dt+2\pi x \int_1^\infty {\pi^2 B_2(t-[t])\over (x^2+(2\pi t)^2)^{{3\over
2}}} dt \ .
\end{equation}
The remainder may be written as a sum,
\begin{equation}
2\pi x \int_1^\infty {\pi^2 B_2(t-[t])\over
(x^2+(2\pi t)^2)^{{3\over 2}}} dt = 2\pi x \sum_{m=1}^\infty \int_0^1
{\pi^2 B_2(t)\over (x^2+(2\pi (t+m))^2)^{{3\over 2}}} dt \ ,
\end{equation}
which may then be expanded for small $x$ in a binomial series;
\begin{equation}
2\pi x \sum_{m=1}^\infty \int_0^1
{\pi^2 B_2(t)\over (x^2+(2\pi (t+m))^2)^{{3\over 2}}} dt =
2\pi^3 x \sum_{k=0}^\infty {-{3\over 2}\choose k} {x^{2k}\over
(2\pi)^{3+2k}}
\sum_{m=1}^\infty \int_0^1 {B_2(t)\over (t+m)^{3+2k}}  dt \ .
\end{equation}
Finally, we rewrite this expression in term of the generalized zeta
function
\begin{equation}
G(x) =2\pi x \int_0^1 {\pi^2 B_2(t)\over (x^2+(2\pi t)^2)^{{3\over 2}}}
dt+2\pi^3 x \sum_{k=0}^\infty {-{3\over 2}\choose k} {x^{2k}\over
(2\pi)^{3+2k}}\int_0^1 B_2(t) \zeta(3+2k,t) \, dt \ .
\end{equation}

All of these integrals are strongly convergent, which is obvious from
simple power counting for all but the $k=0$ term, which we will explicitly
evaluate. The integral in the first term, which is dominant for $x \approx
0$, is elementary
\begin{equation}
\int_0^1 {B_2(t)\over (x^2+(2\pi t)^2)^{{3\over 2}}}={1\over 6x^2
\sqrt{x^2+4\pi^2}}-{1\over 4\pi^2|x|}+{1\over 8\pi^3}\sinh^{-1}
{2\pi\over |x|} \ .
\end{equation}

For the $k=0$ term, we replace the sum of integrals by its original form to
obtain
\begin{equation}
\int_1^\infty {B_2(t-[t])\over t^3} dt={1\over 6}\int_1^\infty {dt\over
t^3} +\lim_{N\rightarrow \infty} \left( -\,\int_1^N {(t-[t])\over t^3} \,
dt+\int_1^N {(t-[t])^2\over t^3} \, dt\right) \ . \label{eq:sumk0}
\end{equation}

We evaluate each of the integrals separately. The first integral is trivial
and gives ${1\over 12}$. Using the notation $F_2(m)={m^{-2}}$, we find that
the second integral is
\begin{eqnarray}
\lim_{N\rightarrow \infty} \int_1^N -\,{(t-[t])\over t^3} \, dt
& = & -\lim_{N\rightarrow \infty}\left(\int_1^N {dt\over t^2}-\sum_{m=1}^{N-1}
m\int_0^1 {dt\over (t+m)^3}\right)\nonumber \\
& = & -\lim_{N\rightarrow \infty} \left(1-{1\over N} +{1\over
2}\sum_{m=1}^{N-1} 
m\left(F_2(m+1)-F_2(m)\right)\right) \nonumber \\ 
& = & -\lim_{N\rightarrow \infty} \left(1-{1\over N} +{1\over
2}\left(-\sum_{m=1}^N F_2(m) +N \, F_2(N)\right)\right)\label{eq:sumbypart1}\\
& = & -\left(1 - {1\over 2}\zeta(2)\right) = {\pi^2\over 12} - 1 \ .
\label{eq:sumk0two} 
\end{eqnarray}

We can evaluate the last integral similarly, this time using the notation
$F_1(m) = m^{-1}$.
\begin{eqnarray}
& & \lim_{N\rightarrow \infty} \int_1^N {(t-[t])^2\over t^3} \, dt =
\lim_{N\rightarrow \infty} \left(\ln N -2\sum_{m=1}^{N-1}\int_0^1 {m \,
dt\over (t+m)^2} +\sum_{m=1}^{N-1}\int_0^1 {m^2 \, dt\over (t+m)^3}\right)
\nonumber \\ &=&\lim_{N\rightarrow \infty}\left( \ln N +2 \sum_{m=1}^{N-1}
m( F_1(m+1)-F_1(m)) -{1\over 2} \sum_{m=1}^{N-1}
m^2(F_2(m+1)-F_2(m))\right) \ .
\end{eqnarray}
Rearranging the series according to Eq.~(\ref{eq:sumbypart1}) and
\begin{equation}
\sum_{m=1}^{N-1} m^2(F_2(m+1)-F_2(m))=-\sum_{m=1}^N (2m-1) F_2(m)+N^2
F_2(N) \ ,
\end{equation}
we obtain
\begin{eqnarray}
\lim_{N\rightarrow \infty} \int_1^N {(t-[t])^2\over t^3} \, dt &=&
\lim_{N\rightarrow \infty}\left( \ln N +2{N\over N}-2\sum_{m=1}^N {1\over
m} -{1\over 2}\left({N^2\over N^2}-\sum_{m=1}^N {(2m-1)\over
m^2}\right)\right)\nonumber  \\ & =
& \lim_{N\rightarrow \infty}\left( \ln N -\sum_{m=1}^N {1\over m} +{3\over
2}-{1\over 2}\sum_{m=1}^N {1\over m^2}\right) \\ & = & {3\over
2}-\gamma-{\pi^2\over 12} \ . \label{eq:sumk0three}
\end{eqnarray}

We put all the terms together from Eqs.~(\ref{eq:sumk0}),
(\ref{eq:sumk0two}), and (\ref{eq:sumk0three}) to find 
\begin{equation}
2\pi^3 x {-{3\over 2}\choose 0} {1\over (2\pi)^3} \int_1^\infty {B_2(t-[t])
\over t^3}\, dt = {x\over 4}\left({7\over 12}-\gamma\right) \ , 
\end{equation}
which implies that 
\begin{equation}
G(x) = \sum_{n=1}^\infty {1\over n} K_1(nx) \approx {\pi^2\over 6 x } -
{\pi \over 2}{\rm sign}(x) + \Big({1-2\gamma\over 8 }\Big) x + {x\over 4}
\sinh^{-1}\left({2\pi\over |x|}\right) + 
{\cal O}(x^3) \ .
\end{equation}

\section{QCD string interacting with a background monopole density}\label{sec:monopole} 

The physics of charged particles moving in a plane under the influence of a
constant magnetic field is a very beautiful and enlightening subject
containing such surprises as the quantum Hall effect.  The string analog of
a charge moving in a constant magnetic field is the motion of a string in a
constant Kalb-Ramond \cite{kalbramond} field strength
\begin{eqnarray}
H_{0ij}(x) &=& 0 \ , \nonumber \\
H_{ijk}(x) &=& h\epsilon_{ijk} \ . \label{eq:Hijk}
\end{eqnarray}
This system is a model of a QCD string interacting with a constant
monopole density and, like the Landau problem, is also exactly solvable
\cite{lundregge}.  (The monopole current is proportional to the dual of
$H_{\alpha\beta\gamma}$: $J^\mu_{\rm monopole} \propto
\epsilon^{\mu\alpha\beta\gamma} H_{\alpha\beta\gamma}$.)  The Kalb-Ramond
potential $B_{\mu\nu}$ corresponding to the field strength (\ref{eq:Hijk}) is
\begin{eqnarray}
B_{0i}(x) &=& 0 \ ,\nonumber \\
B_{ij}(x) &=& {h\over 3} \epsilon_{ijk} x^k \ . \label{eq:Bij}
\end{eqnarray}
In vortex mechanics, the Kalb-Ramond field describes the
Magnus forces on a vortex line moving in a fluid.

The first-quantized action for a bosonic string moving in a background
Kalb-Ramond field is
\begin{equation}
S = -\,{a \over 2}\int d\sigma\, d\tau
\,\left(\sqrt{-h}\, h^{ab} \partial_a X^\mu 
\partial_b X^\nu \eta_{\mu\nu} +  B_{\mu\nu}(X)\epsilon^{ab} \partial_a X^\mu 
\partial_b X^\nu\right) \ .
\end{equation}

If we substitute the field $B_{ij}$ above into the string action, and we
choose orthonormal coordinates on the worldsheet, we find the action becomes
\begin{equation}
S = -\,{a\over 2} \int d\sigma\, d\tau
\,\left(\eta^{ab} \partial_a X^\mu 
\partial_b X^\nu +  {h\over 3} \epsilon_{ijk} \epsilon^{ab} X^i \partial_a X^j 
\partial_b X^k\right) \ ,
\end{equation}
and implies the equations of motion
\begin{eqnarray}
\ddot X^0 - X^{0\prime\prime}&=& 0 \, ,\\
\ddot {\bf X} - {\bf X}^{\prime\prime} &=& -h\dot{\bf X}\times{\bf
X}^\prime \ .
\end{eqnarray}

Here we have used reparametrization invariance of the action to set
coordinate time equal to laboratory time $\tau = x^0$ and to set the
spatial parameter of the string equal to the distance in real space along
the string, $\sigma = z$.  We wish to find the Casimir energy of a long
straight string acting under the influence of this external Kalb-Ramond
field.  We assume the string lies along the $\bf z$ direction and has
length $r$.  Thus we find that $\bf X^\prime = \hat z$, and the linearized
equations of motion for the small vibrations on the string become
\begin{eqnarray}
\ddot X - X^{\prime\prime} &=& -h \dot Y \ ,\nonumber \\
\ddot Y - Y^{\prime\prime} &=& h \dot X \ .
\end{eqnarray}
We define the right and left circularly polarized modes $X^\pm \equiv X \pm
i Y$, and find that in terms of them the small amplitude vibration
equations above reduce to
\begin{equation}
\ddot X^\pm - X^{\pm\prime\prime} = \mp i h \dot X^\pm \ .
\end{equation}

Since our string has fixed length, we know that the wavelength with either
fixed Dirichlet or Neumann boundary conditions (infinitely massive or
completely massless quarks at the end) lead to 
\begin{equation}
\ddot X^\pm \pm i h \dot X^\pm = X^{\pm\prime\prime}= -\left({\pi n\over
r}\right)^2 X^\pm\, ,\quad n=1,2,3,\ldots \ .
\end{equation}
or, if we assume that $X = e^{i\omega t}X_0$,
\begin{equation}
\omega_{\pm}^2 \pm  h \omega_{\pm} = \left({\pi n\over r}\right)^2 = k_n^2
\ .
\end{equation}
The frequencies of small vibrations are then
\begin{equation}
\omega_{n\,\pm} = \mp {h\over 2} +  \sqrt{ k_n^2 + (h/2)^2} = \mp {h\over
2} +  {\pi \over r}\sqrt{ n^2 + \left({r h\over2\pi}\right)^2} \ .\label{KRf}
\end{equation}
Naively, the Casimir energy as a function of the Kalb-Ramond field strength
$h$ is half the sum of all the frequencies
\begin{equation}
E_{\rm Casimir}(h) = {\pi \over r} \sum_{n=1}^\infty \sqrt{ n^2 + \left({r
h\over2\pi}\right)^2} \ .
\end{equation}

Mode sums of this form,
\begin{equation}
\sigma(\xi) = \sum_{n=1}^\infty \sqrt{n^2 + \xi^2} \ ,
\end{equation}
appear in a large variety of problems (see the references following
Eq.~(5.33) in \cite{lambiasenesterenko}).  We now show the simple steps
with which this sum can be expressed in a form very similar to
Eq.~(\ref{Ksum}).

\noindent
First, we subtract out the divergence from the sum in order to 
regularize it
\begin{equation}
\sigma^{\rm reg}(\xi) = \sum_{n=1}^\infty (\sqrt{n^2 +\xi^2} - n) -
\,{1\over12} \ .
\end{equation}
We may rewrite this as
\begin{eqnarray}
\sigma^{\rm reg}(\xi) & = & -\,{1\over12} + \sum_{n=1}^\infty
\int_0^{|\xi|} {x \, dx \over \sqrt{n^2 + x^2} } \\ 
& = & -\,{1\over12} +
\sum_{n=1}^\infty \int_0^{|\xi|} x \, dx \int_0^\infty
{1\over \Gamma({1\over2})}\, t^{-1/2} e^{-\,t(n^2 + x^2)} \, dt \ .
\end{eqnarray}

\noindent
The sum may be done by Poisson resummation
\begin{equation}
\sum_{n=1}^\infty e^{- t n^2}  = {\textstyle 1\over 2}\sqrt{{\pi \over t}} -
{1\over 2} + \sqrt{{\pi \over t}} \sum_{n=1}^\infty e^{-\,{n^2\pi^2\over
t}} \ ,
\end{equation}
leading to
\begin{equation}
\sigma^{\rm reg}(\xi) = -\, {1\over12} + \int_0^{|\xi|} x \, dx
\int_0^\infty {1\over\sqrt{\pi}}\,\, {dt\over\sqrt{t}}\, e^{-\,{t
x^2}}\Big({\textstyle 1\over 2}\sqrt{{\pi \over t}}\, -\, {1\over 2} +
\sqrt{{\pi \over t}} \sum_{n=1}^\infty e^{-\,{n^2\pi^2\over t}}\Big) \ .
\end{equation}

\noindent
Concentrating on the last term, we may use
\begin{equation}
\int_0^\infty x^{\nu -1} \, e^{-\,x \,-\,{\mu^2\over 4x}} \, dx =
2\Big({\mu\over 2}\Big)^\nu K_{-\nu}(\mu) \ ,
\end{equation}
where $K_\nu$ is a modified Bessel function.  Letting $s = t x^2$, we
rewrite the last term in the sum as
\begin{eqnarray}
\int_0^{|\xi|} x \, dx \int_0^\infty
{1\over\sqrt{\pi}}\,\, {dt\over\sqrt{t}} e^{-{t x^2}}\Big(\sqrt{{\pi \over t}}
\sum_{n=1}^\infty e^{-\,{n^2\pi^2\over t}}\Big) 
&=& 
\sum_{n=1}^\infty \int_0^{|\xi|} x \, dx \int_0^\infty {ds\over s}
e^{-s\,-\,{n^2\pi^2 x^2\over s}} \nonumber \\  
&=& 2 \sum_{n=1}^\infty \int_0^{|\xi|} x \, dx\,  K_0(2 n \pi x) \nonumber \\
&=&{1\over12} - \sum_{n=1}^\infty {|\xi|\over n\pi} K_1(2 n \pi\, |\xi|) \ .
\end{eqnarray}
Finally, we have 
\begin{equation}
\sigma^{\rm reg}(\xi) = -\, {|\xi|\over \pi}\sum_{n=1}^\infty {K_1(2 n
\pi\,|\xi|)\over n} + {1\over 2\sqrt{\pi} } \int_0^{|\xi|} x \, dx
\int_0^\infty t^{-1/2} \Big(\sqrt{{\pi\over t}}  - 1 \Big)e^{-x^2 t}\, dt \ ,
\end{equation}
in which the final integral may be expanded and evaluated in terms of Gamma
functions as
\begin{eqnarray}
{1\over 2\sqrt{\pi} } \int_0^{|\xi|} x \, dx
\int_0^\infty t^{-1/2} \Big(\sqrt{{\pi\over t}}  - 1 \Big)e^{-x^2 t}\, dt &=&
{1\over 2\sqrt{\pi} } \int_0^\infty t^{-1/2} \Big(\sqrt{{\pi\over t}}  - 1
\Big) {1\over 2t} \Big(1 - e^{-\xi^2 t}\Big) \, dt \nonumber \\
&=& {1\over 4 \sqrt{\pi} } \int_0^\infty \Big( \sqrt{{\pi\over t}} -
\sqrt{{\pi\over t}} e^{-\xi^2 t} - 1 + e^{-\xi^2 t} \Big) \,
{dt\over t^{3/2}} \nonumber \\ 
&=& \infty + {1\over 4 \sqrt{\pi} } \int_0^\infty e^{-\xi^2 t}
t^{-3/2} \, dt  \nonumber \\ 
&=& \infty + \infty\,\, \xi^2 - {|\xi|\over 2} \ .  
\end{eqnarray}
Subtracting the infinite pieces above, we find the final regularized result
\begin{equation}
\sigma^{\rm reg}(\xi) = -\, {|\xi|\over 2}\, -\,{ |\xi| \over \pi}
\sum_{n=1}^\infty {K_1(2 n \pi\,|\xi|)\over n} = -\, {|\xi|\over 2}\, -\,{
|\xi| \over \pi}G(2\pi\,|\xi|) \ .
\end{equation}
This and related results are found by a slightly different method in
Ref.~\cite{nesterenkopirozhenko}.

The Casimir energy corresponding to the zero-point frequencies (\ref{KRf}), 
\begin{equation}
E_{\rm Casimir} = {\pi\over r}\,\sigma({hr/2\pi}) \ ,
\end{equation}
can be evaluated using the results of section \ref{sec:Besselsum}.  We find
\begin{equation}
E_{\rm Casimir} =  -\, {\pi\over 12 r}  + {h^2 r\over
16\pi}\left[(2\gamma -1)  - 2 \sinh^{-1}\left({2\pi\over |h|r}  \right)
\right]\, + {\cal O}(h^3 r^2) \ . 
\end{equation}
For $h\rightarrow 0$ this reduces to the fixed-end straight string result
for two transverse degrees of freedom.

\section{Discrete Isgur-Paton string}\label{sec:discrete}

In this section we consider the interquark potential of a static
quark-antiquark pair bound by a flux tube, considered by Isgur and Paton
\cite{isgurpaton} and Merlin and Paton \cite{merlinpaton} to be a discrete
string consisting of $N$ masses $m$ a distance $d$ apart.  This string,
when terminated by particles of mass $M$, has equations of motion
\cite{isgurpaton,merlinpaton,allenschmidt} with the secular
equation
\begin{eqnarray} 
1-{M\over m}\lambda &=& {\sin(N\theta)\over \sin((N+1)\theta)} \ , \\
\lambda &=& 2(1-\cos\theta)\ , \qquad \lambda={\omega^2\over \beta} \ ,
\end{eqnarray}
for the transverse vibrational frequencies $\omega$. Here $\beta$ is
defined in terms of the masses $m$, the bare tension $a_0$ and interquark
distance $r$, as $\beta = {a_0 (N+1)\over mr}$. The
infinite mass $M$ limit results in
\begin{equation}
\theta={p\pi\over (N+1)} \ , \qquad p=1,2,\dots , N \ .
\end{equation}
 Assuming finite $M$ corrections of the form
\begin{equation} 
\theta\rightarrow {p\pi+\delta\over (N+1)} \ ,
\end{equation}
we discover that
\begin{equation}\label{38} 
\omega \approx \sqrt{\beta} \left( \sin\left({p\pi\over 2(N+1)}\right) +
{m\over 4M(N+1)} \left({1\over \sin\left({p\pi\over 2(N+1)}\right)} -
\sin\left({p\pi\over 2(N+1)}\right)\right)\right) \ .
\end{equation} 

\noindent
The contribution of the first and third terms to the mode sum is 
simple, following from
\begin{equation}
\sum_{p=1}^N \sqrt{\beta} \sin\left({p\pi\over 2(N+1)}\right) =
\sqrt{{\beta\over 2}}\,\, {\sin\left({N\pi\over 4(N+1)}\right)\over
\sin\left({\pi\over 4(N+1)}\right)} \ .
\end{equation}
The second term can be analyzed by use of the expansion
\begin{equation} 
\csc(x)={1\over x} +2 \sum_{k=1}^\infty {(2^{2k-1} -1)\over (2k)!}\, |B_{2k}|
\,x^{2k-1} \ ,
\end{equation}
in which $B_k$ are the Bernoulli numbers. \\

\noindent
We begin with
\begin{eqnarray}\label{40} 
J(N) & = & \sum_{p=1}^N \csc\left({\pi p\over 2(N+1)}\right)\nonumber\\
&=&{2(N+1)\over \pi}\sum_{p=1}^N {1\over p} + \sum_{k=1}^\infty 2\,
{(2^{2k-1} -1)\over (2k)!}\, |B_{2k}|\,\left({\pi\over
2(N+1)}\right)^{2k-1} \sum_{p=1}^N p^{2k-1} \ ,
\end{eqnarray}
and use
\begin{equation} 
\sum_{p=1}^N p^{2k-1}={B_{2k}(N+1)-B_{2k}\over 2k} \ ,
\end{equation}
in which the Bernoulli polynomials $B_n(x)$ satisfy the formal recursion
\begin{equation} 
B_n(x)=(B+x)^n\quad{\rm with}\quad B^n\rightarrow B_n \ .
\end{equation}
We find that
\begin{eqnarray} 
\sum_{p=1}^N p^{2k-1}&=&{1\over 2k}(N+1)^{2k}-{1\over 2}(N+1)^{2k-1} +
 {2k-1\over 12}(N+1)^{2k-2}\nonumber \\&\phantom{=} & -\ {(2k-1)(2k-2)(2k-3)\over 30\cdot
 4!}(N+1)^{2k-4}+\cdots \ .
\end{eqnarray}
The second term in Eq.~(\ref{38}) thus contributes five terms to the mode sum
\begin{equation}
J(N)= \sum_{p=1}^N \csc\left({\pi p\over 2(N+1)}\right)=I_0+I_1+I_2+I_3+I_4
\end{equation}
of leading orders in $N$. The first is simple 
\begin{eqnarray}
I_0 &=& {2(N+1)\over \pi}\sum_{p=1}^N {1\over p} \nonumber \\
 &=& {2(N+1)\over\pi}\left(\gamma+\ln N+{1\over 2N}\,
 -\sum_{k=2}^\infty {A_k\over N(N+1)(N+2)\cdots(N+k-1)}\right) \ ,
\end{eqnarray}
for which $A_2=A_3={1\over 12}$, $A_4={19\over 80}$, and so forth. This is
a standard expansion of the Digamma function \cite{GR}. 

\begin{figure}[htbp]
\epsfxsize = \textwidth
\epsfbox{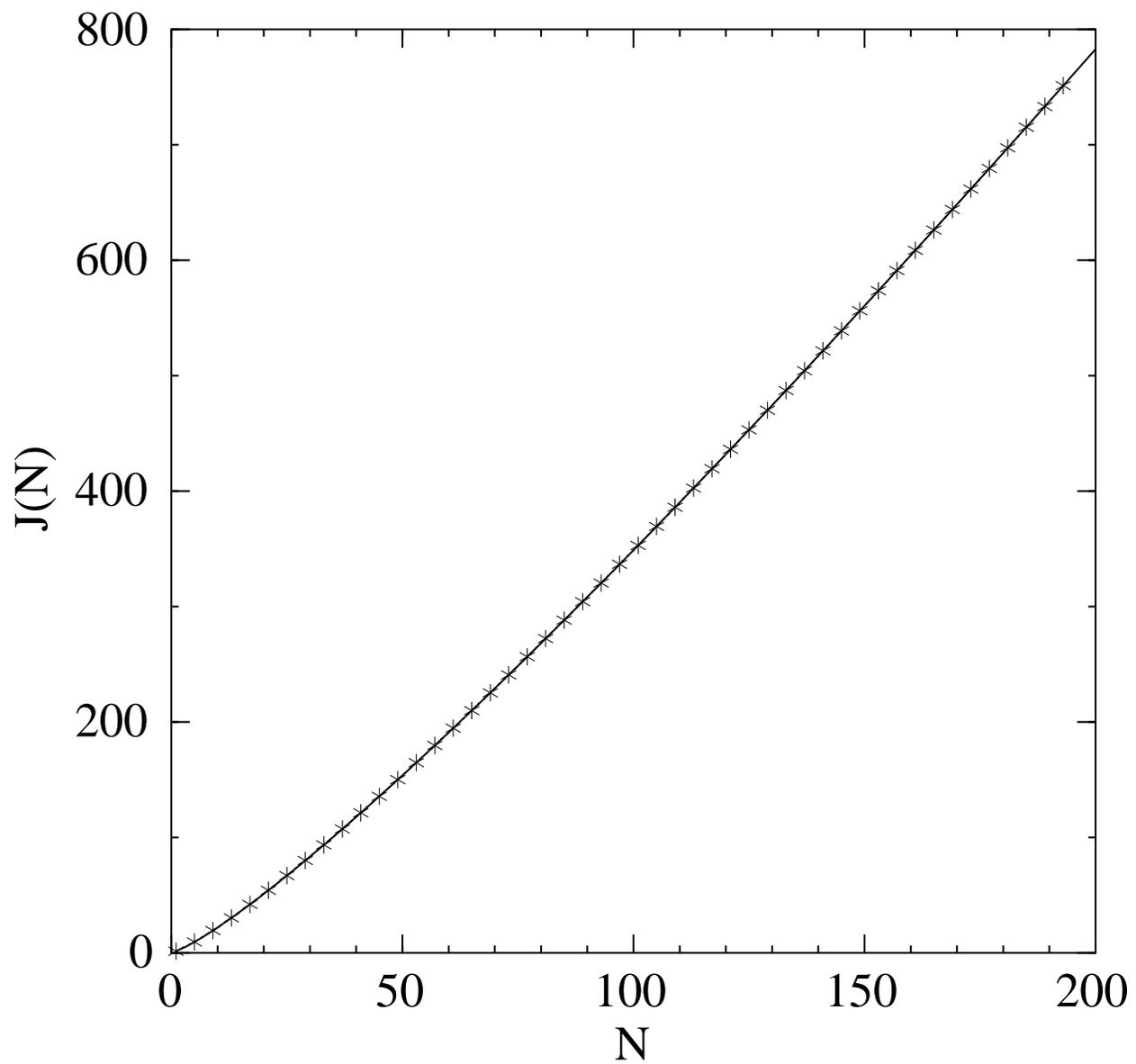}
\vskip 1 cm
\caption{Numerical evaluations of the sum $J(N)$ . The solid curve is the 
function Eq.~(\protect{\ref{54}}), while the diamonds are for the sum 
Eq.~(\protect{\ref{40}}).}
\label{fig:three}
\end{figure}

The next term is
\begin{eqnarray} 
I_1&=& {2(N+1)\over \pi}\sum_{k=1}^\infty 2\,{(2^{2k-1} -1)\over (2k)!}\,
|B_{2k}|\, {1\over 2k}\left({\pi\over 2}\right)^{2k} \\
&=&{2(N+1)\over\pi}\int_0^{{\pi\over 2}}\left(\csc x-{1\over x}\right)
dx={2(N+1)\over\pi}\left(\ln{2\over \pi}-\ln{1\over 2}\right) \ ,
\end{eqnarray}
and subsequent terms can all be expressed as valuations of the derivatives
of $\csc(x)-{1\over x}$ at the point $x={\pi\over 2}$.
\begin{eqnarray}
I_2 &=& -\,{1\over 2}\sum_{k=1}^\infty 2{(2^{2k-1} -1)\over (2k)!}\,
|B_{2k}|\, \left({\pi\over 2}\right)^{2k-1} \\ &=&-\,{1\over 2}\left(\csc
x-{1\over x}\right)\Bigg|_{x={\pi\over 2}}={1\over\pi}-{1\over 2} \ , \\ I_3
&=& {1\over 6(N+1)}\,\sum_{k=1}^\infty 2\,{(2^{2k-1} -1)\over (2k)!}\,
|B_{2k}|\, {2k-1\over 2}\, \left({\pi\over 2}\right)^{2k} \\ &=&{1\over
12(N+1)}\,\, x\,{d\over dx} \left(\csc x-{1\over
x}\right)\Bigg|_{x={\pi\over 2}} ={\pi\over 24(N+1)} \ ,
\end{eqnarray}
and finally
\begin{equation}
I_4=-\,{1\over 30\cdot 4!\, (N+1)^3}\,\, x^3\,{d^3\over dx^3} \left(\csc
x-{1\over x}\right)\Bigg|_{x={\pi\over 2}} \ .
\end{equation}
All of the parts can be assembled to give
\begin{eqnarray}\label{54}
J(N)=\sum_{p=1}^N \csc\left({\pi p\over 2(N+1)}\right)&=& {2(N+1)\over
\pi}\left(\gamma+\ln{4(N+1)\over\pi}\right) - \, {1\over 2}\nonumber \\
&+&{\pi\over 24(N+1)} + \left({1\over\pi}-{1\over 12}\right){1\over
N(N+1)}+ {\cal O}({N^{-3}}) \ .
\end{eqnarray}
This expression agrees very well with the actual sum, plotted as a function
of $N$. This is illustrated in Fig.~\ref{fig:three}.

For large quark masses $M$, we can again calculate the lowest order $1\over
M$ corrections to the Casimir energy, this time for the discrete string
before the limit of link spacing going to zero ($d\rightarrow 0$) is taken.
To make contact with Isgur and Paton \cite{isgurpaton}, we take $\beta =
{1\over d^2}$ and then use $N = ({r\over d}) - 1$ to arrive at the Casimir
energy contribution per transverse degree of freedom
\begin{equation}\label{discreteCasimir}
E_{\rm Casimir} = {2 r\over \pi d^2} - {1\over 2 d} - {\pi \over 24 r}
- {a_0 (1-\gamma -\ln({4r\over \pi d}))\over 2\pi M} + {\cal O}({1\over
M^2})\ .
\end{equation}

We note that although the universal L\"uscher term $-\,{\pi \over 24 r}$ is
present in both (\ref{continCasimir}) and (\ref{discreteCasimir}), the 
$1\over M$ corrections are completely different.  In the former case,
the correction is ${\cal O}({1\over\sqrt{M}})$, while in the latter it is
${\cal O}({1\over{M}})$.  Of course, there is no reason to expect that the
form of the Casimir energies will be the same for the Nambu-Goto string and
the discrete string of Isgur and Paton.  The former is relativistic while
the later is not.  In addition, it is not likely that the Casimir energy
of a discrete string will go the the continuum result as the number of mass
points is taken to infinity.  Sums and limits generally depend upon the
order in which they are done, especially when the sums are divergent and
must be regularized.

\section{Conclusions}
We have shown how to regularize and perform the zero-point energy mode sums
for the discrete non-relativistic string studied by Isgur and Paton and the
relativistic continuum Nambu-Goto string. Computation of these divergent
sums is necessary in order to obtain the Casimir energy of the system to be
interpreted as part of the static potential for quarks bound by a QCD
string.  We have shown that the lowest order corrections to the Casimir
energies in the reciprocal of the mass of the terminating quark is
different for the two cases.  We have also shown how the addition of a
background Kalb-Ramond field with a corresponding constant field strength
affects the static potential.  We hope to extend these results to the case
of a string interacting with itself through a Kalb-Ramond field.

\acknowledgments 

We thank S. Dowker, R. Easther, and V.V. Nesterenko for useful
comments on an earlier version of this paper.  This work was supported in
part by the US Department of Energy under Contract No.~DE-FG02-95ER40896.


\vspace{2cm}
\newpage
\vspace{2cm}

\vspace{2cm}
\newpage
\vspace{2cm}

\end{document}